\documentclass[fleqn,usenatbib,useAMS]{mnras}

\usepackage{orcidlink}
\usepackage{newtxtext,newtxmath}
\usepackage[T1]{fontenc}
\DeclareRobustCommand{\VAN}[3]{#2}
\let\VANthebibliography\thebibliography
\def\thebibliography{\DeclareRobustCommand{\VAN}[3]{##3}\VANthebibliography}

\usepackage{xcolor} 

\newcommand{\Fermi}{{\itshape Fermi}}
\newcommand{\FermiGBM}{\Fermi\//GBM}
\newcommand{\Swift}{{\itshape Swift}}
\newcommand{\SwiftBAT}{\Swift\//BAT}
\newcommand{\sqdeg}{\mbox{$\mbox{deg}^2$}}

\usepackage{graphicx}	
\usepackage{amsmath}	
\usepackage{comment}

\usepackage{pdflscape}

\newcommand{\CHANGE}[2]{{ #2}}
\newcommand{\ADD}[1]{{ #1}}

\title[BALROG and RoboBA]{An evaluation of the BALROG and RoboBA algorithms for determining the position of {\FermiGBM} GRBs}

\author[K.\ O.\ C.\ López et al.]{K.~Ocelotl.~C.~López\,\orcidlink{0000-0002-9322-6900},$^{1}$\thanks{E-mail: koclopez@astro.unam.mx (KOCL)}
Alan~M.~Watson\,\orcidlink{0000-0002-2008-6927},$^{1}$
William~H.~Lee\,\orcidlink{0000-0002-2467-5673},$^{1}$
Rosa~L.~Becerra\,\orcidlink{0000-0002-0216-3415},$^{2,3}$
\newauthor
and
Margarita~Pereyra\orcidlink{0000-0001-6148-6532}$^{4}$
\\
$^{1}$Instituto de Astronomía, Universidad Nacional Autónoma de México, Apartado Postal 70-264, 04510 México, CDMX, Mexico\\
$^{2}$Instituto de Ciencias Nucleares, Universidad Nacional Autónoma de México, Apartado Postal 70-264, 04510 México, CDMX, Mexico\\
$^{3}$Department of Physics, University of Rome - Tor Vergata, via della Ricerca Scientifica 1, 00100 Rome, IT\\
$^{4}$CONAHCyT, Instituto de Astronomía, Universidad Nacional Autónoma de México, 22860 Ensenada, BC, México
}

\date{Accepted XXX. Received YYY; in original form ZZZ}

\pubyear{2024}

\begin{document}
\label{firstpage}
\pagerange{\pageref{firstpage}--\pageref{lastpage}}

\maketitle

\begin{abstract}
The {\FermiGBM} instrument is a vital source of detections of gamma-ray bursts and has an increasingly important role to play in understanding gravitational-wave transients. In both cases, its impact is increased by accurate positions with reliable uncertainties.
We evaluate the RoboBA and BALROG algorithms for determining the position of gamma-ray bursts detected by the {\FermiGBM} instrument. We construct a sample of 54 bursts with detections both by {\SwiftBAT} and by {\FermiGBM}. We then compare the positions predicted by RoboBA and BALROG with the positions measured by BAT, which we can assume to be the true position. We find that RoboBA and BALROG are similarly precise for bright bursts whose uncertainties are dominated by systematic errors, but RoboBA performs better for faint bursts whose uncertainties are dominated by statistical noise. We further find that the uncertainties in the positions predicted by RoboBA are consistent with the distribution of position errors, whereas BALROG seems to be underestimating the uncertainties by a factor of about two. \CHANGE{We further}{Additionally, we} \CHANGE{evaluate the strategy for the follow-up of the optical afterglows of {\FermiGBM} bursts. with the DDOTI wide-field imager, and}{consider the implications of these results for the follow-up of the optical afterglows of {\FermiGBM} bursts. In particular, for the DDOTI wide-field imager we} conclude that a single pointing is best. Our sample would allow a similar study to be carried out for other telescopes.


\end{abstract}


\section{Introduction}
\label{sec:introduction}

Gamma-ray bursts (GRBs) are transient events with isotropic energies of $10^{50}$--$10^{53}$ ergs caused by a collimated relativistic jet launched from the vicinity of a compact central engine \citep{gehrels2013gamma}. They consist of an initial prompt emission of gamma-ray photons, largely from internal shocks in the jet, followed by an afterglow at longer wavelengths resulting from the interaction of the jet with the circumstellar medium \citep{piran1998spectra,granot2002shape}. 

\CHANGE{We can classify GRBs}{GRBs are empirically classified} into two populations using the time interval $T_{90}$ between the moments at which $5\%$ and $95\%$ of the prompt emission are detected \citep{kouveliotou1993identification}. GRBs are classified as short (SGRB) if they have a $T_{90}<2$~s and long (LGRB) if they have a $T_{90}>2$~s. This classification is important as there is a strong correlation between $T_{90}$ and the progenitor system. LGRBs \CHANGE{appear to result from}{tend to be the result of} the death of massive stars following core collapse \citep{woosley1993gamma,woosley2006supernova}, whereas SGRB \ADD{tend to} arise from the merger of compact binary systems with at least one neutron star \citep{lee2007progenitors,abbott2017gravitational,eichler1989nucleosynthesis,narayan1992gamma,ruffert1998colliding}. \CHANGE{SGRBs}{These mergers} are caused by the orbital decay of the compact binary system due to the emission of gravitational radiation, and as such are important for multi-messenger astronomy \citep{branchesi2016multi}. The two populations do overlap to some degree, and so there is ambiguity when $T_{90}$ is around 2~s \citep{dimple2022grb,garcia2024identification,yang2022long,troja2022nearby}.

For many reasons, the afterglow phase is vitally important for the identification and subsequent follow-up of GRBs. Gamma-ray detectors tend to have poor angular resolution, so observations of the afterglow in X-rays or the optical are the best means to \ADD{obtain a localization at the level of arcseconds. Such localizations allow us to} tie a GRB to its host galaxy or conversely demonstrate that it occurs outside of a galaxy. \CHANGE{All redshift measurements}{They are also necessary for redshift determination, which} to date have been made either directly from the afterglow or from the host galaxy associated with the afterglow. Finally, the afterglow provides important information on the environment in which the GRB occurred.

The {\Swift} satellite \citep{gehrels2004swift} has allowed us to exploit these characteristics of GRBs and their afterglows over the last 19 years. GRBs are detected in gamma rays by the BAT instrument \citep{barthelmy2005bat}, and then the afterglow is localized to arcsec-precision in X-rays by the XRT \citep{burrows2005xrt,evans2009xrt} and in the optical by UVOT \citep{roming2005uvot,roming2017large,page2019study}. The main disadvantages of {\Swift} are the relatively small field-of-view of BAT of about 1.4~sr (only about 11\% of the sky), the relatively soft response of BAT (up to about 150~keV), and worries about its longevity given its age and reliance on reaction wheels \citep{cenko2022reactionwheel}.

The {\Fermi} satellite has several advantages compared to {\Swift}. Its GBM instrument views 70\% of the sky and has good sensitivity up to 10~MeV \citep{meegan2009fermi}. These properties allow it to detect about 250 GRBs per year \citep{vonkienlin2020catalog}, compared to about 100 for BAT \citep{lien2016catalog}. Furthermore, its response at higher energies allows it to detect about 40 SGRBs per year \citep{kocevski2018sgrb}, whereas BAT detects only about 10 \citep{liensakamoto2014}. On the other hand, the localizations delivered by GBM have typical uncertainties of 5--15~deg \citep{connaughton2015localization}, and {\Fermi} is not equipped with a means to improve this by observing emission from the afterglow. Some GBM GRBs are also detected at higher energies by the LAT instrument and localized to about 10~arcmin, but only about 20 per year \citep{ajello2019ApJcatalog}. 

Instead, precise localizations of GBM GRBs must typically be provided by the detection of the optical afterglow by wide-field imagers \CHANGE{ \citep{singer2015phd,watson2016ddoti,mong2021goto,ahumada2022search}}{such as iPTF \citep{singer2015phd}, DDOTI \citep{,watson2016ddoti}, GOTO \citep{mong2021goto}, and ZTF \citep{ahumada2022search}}. The success of these searches is improved by having not only good estimates of the positions from GBM but also good estimates of the uncertainties in these positions.

Moreover, GBM has acquired new importance in the era of gravitational-wave astronomy, as its wide field gives an excellent chance of detecting faint gamma-ray emission associated with nearby gravitational-wave sources such as compact binary mergers. This was dramatically demonstrated in the case of GRB~170817A, which was produced by GW170817 \citep{abbott2017gravitational,goldstein2017gw,savachenko2017gw}. Detections by GBM allow the search area to be narrowed and provide vital information on the nature of the progenitors and remnant. \ADD{One such a system is RAVEN adopted by the LVK collaboration \citep{Adhikari2023A, Chaudhary2023}.}

For these reasons, accurate positions and reliable uncertainties for GBM detections are increasingly important. In this work, we evaluate two current systems that estimate the position of GBM GRBs: the RoboBA system \citep{connaughton2015localization,goldstein2020evaluation} and the BALROG system \citep{burgess2018,berlato2019improved}. In contrast to previous work, we will use published positions produced by the teams behind both systems. Thus, our evaluation is entirely empirical and independent.

Our paper is organized as follows. In section \ref{section:gbm} we briefly describe the GBM instrument and the means by which it provides information for localizing GRBs. In section \ref{section:positions} we describe the products of the RoboBA and BALROG processes. In section \ref{section:samples} we describe the samples we use to evaluate the performance of RoboBA and BALROG. In section \ref{section:analysis} we consider the accuracy of both the position estimates and the uncertainty estimates. In section \ref{section:ddoti} we determine the most appropriate observation strategy for our wide-field imager DDOTI based on our results. Finally, in section \ref{section:discussion} we summarize and discuss our results.

\section{GBM Positions}
\label{section:gbm}

{\FermiGBM} has twelve sodium iodide (NaI) detectors sensitive over an energy range from 8 keV to 1 MeV and two bismuth germanate (BGO) scintillators sensitive from 200~keV to 40~MeV \citep{meegan2009fermi}. The NaI detectors are distributed around the spacecraft and oriented in different directions. The two BGO detectors are on opposite sides of the spacecraft. The signal measured in each detector depends on the position and spectrum of the source for two main reasons: each detector has an energy-dependent angular response function, and absorption in the spacecraft reduces the count rate in detectors on the side that faces away from the source. In addition, the scattering of gamma rays from the Earth's atmosphere also changes the relative count rates. By modeling the count rates, taking into account the spectrum of the source, the position can be determined, albeit with significant uncertainties.

The {\Fermi} spacecraft provides initial “flight” positions calculated by an on-board computer. While these are produced in 10––30~s \citep{connaughton2015localization}, the meager processing power available limits their accuracy. Better “ground” positions are provided by subsequent processing of downloaded data using more powerful computers on the Earth.

In the first years of the {\Fermi} mission, ground positions were provided by having the Burst Advocates manually run the Daughter of Location (DoL) algorithm \citep{connaughton2015localization}. This algorithm fits the count rates by varying the position of the source and allowing it to have one of three different spectra representing soft, medium, and hard GRBs. In their evaluation of DoL, \cite{connaughton2015localization} found that the statistical uncertainties for bright GRBs could be as small as 1~deg, but the systematic uncertainties were well represented by a Gaussian with $1\sigma$ radius of 3.7~degrees and a non-Gaussian tail containing about 10\% of the probability and extending to approximately 14~deg.

The BALROG algorithm was created by \cite{burgess2018}. The major advance was noting that there were correlations between the position and spectrum of the GRB, and so in theory a better determination of the uncertainties could be obtained by fitting simultaneously for both. This could also reduce the apparent systematic error in the determination of the position of bright GRBs. The original BALROG algorithm was improved and automated by \cite{berlato2019improved}, and produces positions in a matter of minutes, although the delay before the GBM data is publically released is an important factor. The BALROG algorithm is specifically tuned for bright GRBs; for example, it uses only about 10~seconds of data, to avoid smearing in the response caused by the motion of the spacecraft. \cite{berlato2019improved} also compared the BALROG results to DoL and found that BALROG gave more precise positions, at least for bright GRBs in which the DoL uncertainty is dominated by systematic effects.

Subsequently, the RoboBA system was deployed. \cite{goldstein2020evaluation} describe it as an automated system that uses an improved version of the earlier Daughter of Location (DoL) and report that the systematic uncertainty for the updated RoboBA localizations was 1.8~degrees for 52\% of GRBs and 4.1 for the remaining 48\%. RoboBA also runs in a matter of minutes and has the advantage of early access to proprietary GBM data. \cite{goldstein2020evaluation} compared the RoboBA results to BALROG, and found that RoboBA gave more precise positions.

\ADD{Both systems distribute their results in a form that is convenient for robotic telescopes. The GBM flight positions and RoboBA positions are distributed using GCN Notices \citep{Barthelmy1998AAS}. The BALROG positions are distributed as JSON and FITS files whose locations can be derived from the GBM trigger number. In this sense, there is little to choose between the two. (The BALROG team also send GCN Circulars, but it is more difficult for a robotic system to automatically extract information from the text in these.)}

Our direct interest in GBM positions is to use them to point our wide-field imager DDOTI \citep{watson2016ddoti}. \ADD{In the first few minutes after a burst we only have GBM flight positions and RoboBA positions. After this initial period, we have both BALROG and RoboBA positions, and obviously we wish to use the better one. This requires understanding the performance of each algorithm.} The situation we faced at the start of this work was that each team had published results that indicated that their approach gave better positions. Therefore, we embarked on this independent empirical study to gain a clearer understanding of the matter.

\section{Position Estimations and Uncertainties}
\label{section:positions}

In this section, we briefly describe the products of the RoboBA and BALROG algorithms and in particular their model uncertainties. This is useful to establish our notation.

Each algorithm provides estimators $\hat\alpha$ and $\hat\delta$ of the true right ascension $\alpha$ and declination $\delta$ of the burst. We will use $\gamma$ to be the total angular distance error (i.e., the angular distance between the estimated position and the true position) and $\gamma_\alpha$ and $\gamma_\delta$ to be the corresponding angular distances parallel to the local right ascension and declination axes.

One potentially confusing aspect is that the error distributions are traditionally described in terms of the circle or ellipse corresponding to $1\sigma$, $2\sigma$, or $3\sigma$ \citep{briggs1999error,connaughton2015localization,berlato2019improved}. This does not mean that the angular radius of the circle is that number of standard deviations. Rather, it means that the probability within the circle or ellipse is the same as that within $\pm1\sigma$, $\pm2\sigma$, and $\pm3\sigma$ of a one-dimensional Gaussian distribution, that is, 0.6827, 0.9545, and 0.9973.

For a two-dimensional circular Gaussian distribution with standard deviation $a$ in each coordinate, the probability density is
\begin{equation}
p(\gamma) = \frac{1}{2\pi a^2} e^{-\gamma^2/2a^2}.
\label{equation:gaussian}
\end{equation}
We can easily integrate this \CHANGE{to obtain the total probability within a radius $\gamma$,
\begin{equation}
P(\le \gamma) = 1 - e^{-\gamma^2/2a^2}
\end{equation}
and then obtain
\begin{equation}
\gamma = a (-2\ln (1-P))^{1/2}.
\end{equation}
We see}{and find} that the $1\sigma$, $2\sigma$, and $3\sigma$ radii correspond to $\gamma = 1.515a$, $2.486a$, and $3.439a$.

\subsection{RoboBA}
\label{section:roboba-definitions}

RoboBA uses the “DoL” algorithm described in detail by \cite{connaughton2015localization} and \cite{goldstein2020evaluation}.

The model error distribution is based on the von~Mises-Fisher distribution \citep{fisher1993statistical,briggs1999error,connaughton2015localization}, which is a generalization of the Gaussian distribution to the surface of a sphere, and is given by
\begin{equation} 
p(\gamma) = 
\frac{\kappa}{2\pi \left ( e^{\kappa}-e^{-\kappa} \right )} e^{\kappa \cos \gamma },
\end{equation}
in which the concentration parameter $\kappa$ is used to characterize the width of the distribution. 
The probability within an angular radius $\gamma$ is
\begin{equation}
    P(\le \gamma) = \frac{\kappa}{2\pi \left ( 1-e^{-2\kappa} \right )} \int_{\Omega} e^{\kappa (\cos \gamma - 1)}
    \,d\Omega.
\end{equation}
Unfortunately, this integral does not, in general, have a closed form.
Therefore, to advance, we rewrite the distribution as 
\begin{equation}
p(\gamma) = \frac{\kappa}{2\pi(1-e^{-2\kappa})} e^{\kappa(\cos\gamma-1)}.
\label{equation:von-mises-fisher}
\end{equation}
This form has two advantages in our current context in which $\gamma$ is typically small and $\kappa$ is typically large. First, this form avoids overflow when evaluated numerically. Second, we can approximate $\cos\gamma$ as $1-\gamma^2/2$ with an error of less than 1 part in 1000 out to a radius of 20~deg, and substituting this into equation~(\ref{equation:von-mises-fisher}) we rapidly obtain an approximate Gaussian distribution,
\begin{equation}
    p(\gamma) \approx \frac{\kappa}{2\pi} e^{-\kappa \gamma^2/2}.
\label{equation:gaussian-from-fisher}
\end{equation}
We can then use all of the standard results for a Gaussian.
Comparing equations \ref{equation:gaussian} and \ref{equation:gaussian-from-fisher}, we see that $\kappa^{-1} \approx a^2$ and so in terms of the $1\sigma$ angular radius $\sigma$ \ADD{in radians} \citep{briggs1999error},
\begin{equation}
\kappa^{-1} \approx 0.660^2\sigma^2.
\end{equation}
The numerical factor here is simply the inverse of the factor 1.515 in the relation between $\sigma$ and $a$ derived above. When $\sigma \le 10$~deg \ADD{(i.e., $\sigma \le 0.18$~rad)}, $\kappa \ge 75$, which validates our assumption that $\kappa$ is typically large.

The DoL algorithm uses two von~Mises-Fisher distributions, one for the core (containing a fraction $f$ of the probability), and one for the tail \citep{connaughton2015localization,goldstein2020evaluation}. That is, the probability that the error is less than the angular distance $\gamma$ is given by
\begin{equation}
P(\le\gamma) = f P_\mathrm{core}(\le\gamma) + (1-f)P_\mathrm{tail}(\le \gamma).
\label{equation:roboba-enclosed}
\end{equation}
The two distributions $P_\mathrm{core}$ and $P_\mathrm{tail}$ are von~Mises-Fisher distributions with different values of the concentration parameter $\kappa_\mathrm{core}$ and $\kappa_\mathrm{tail}$. Each combines the statistical uncertainty $\sigma_\mathrm{stat}$ with different values of the systematic uncertainty $\sigma_\mathrm{sys}$. For $P_\mathrm{core}$, we have
\begin{equation}
\kappa_{\mathrm{core}}^{-1} = 0.660^2(\sigma_\mathrm{stat}^2 + \sigma_\mathrm{core}^2),
\end{equation}
whereas for $P_\mathrm{tail}$, we have
\begin{equation}
\kappa_{\mathrm{tail}}^{-1} = 0.660^2(\sigma_\mathrm{stat}^2 + \sigma_\mathrm{tail}^2).
\end{equation}

The statistical uncertainty $\sigma_\mathrm{stat}$ is distributed with the predicted position. To calculate the model error distribution, we also need the fraction of probability in the core and the systematic uncertainties for the two components. We adopt the “Updated RoboBA All GRBs” model of \cite{goldstein2020evaluation}, which has $f = 0.517$, $\sigma_\mathrm{core} = 1.81~\mathrm{deg}$, and $\sigma_\mathrm{tail} = 4.07~\mathrm{deg}$.

\subsection{BALROG}
\label{section:balrog-definitions}

The BALROG algorithm produces two-dimensional images of the probability distribution of the position on the sky \citep{burgess2018}. Two images are produced for each localization, one with just the statistical uncertainty and another convolved with a two-dimensional Gaussian representing the systematic uncertainty \citep{berlato2019improved}. \ADD{The BALROG team view these images as their primary products (Greiner, private communiation).}

Subsequently, \CHANGE{the unconvolved image is fitted}{the BALROG process fits the unconvolved image} with a two-dimensional Gaussian, 
\begin{equation}
p(\gamma_\alpha,\gamma_\delta) = \frac{1}{2\pi a_\alpha a_\delta} e^{-(\gamma_\alpha^2/a_\alpha^2+\gamma_\delta^2/a_\delta^2)/2}.
\end{equation}
in which $\gamma_\alpha$ and $\gamma_\delta$ are the angular separations in $\alpha$ and $\delta$ and $a_\alpha$ and $a_\delta$ are the corresponding standard deviations \citep{berlato2019improved}. The results of the fit are given as $\sigma_\alpha$ and $\sigma_\delta$, the half-axes of the ellipse that contains 0.683 of the probability. Note that $\sigma_\delta$ is given in terms of $\alpha$, not the separation $\gamma_\alpha$, and so needs to be multiplied by $\cos\hat\delta$ before being used with angular separations. The systematic uncertainty $\sigma_\mathrm{sys}$ is also given as the radius enclosing 0.683 of the probability and is typically 1.0 or 2.0~deg. \ADD{All of these parameters are distributed as secondary products by the BALROG team.}

To obtain an approximation of the parameters $\sigma_\alpha'$ and $\sigma_\delta'$ of an equivalent fit to the image after convolution with the systematic uncertainty, we add the systematic uncertainty in quadrature as follows,
\begin{equation}
    (\sigma_\alpha'\cos\delta)^2 = (\sigma_\alpha\cos\delta)^2 + (\sigma_\mathrm{sys})^2
\end{equation}
and
\begin{equation}
    (\sigma_\delta')^2 = (\sigma_\delta)^2 + (\sigma_\mathrm{sys})^2.
\end{equation}
The final Gaussian then has
\begin{equation}
a_\alpha = 0.660 \sigma_\alpha'\cos\delta
\end{equation}
and
\begin{equation}
a_\delta = 0.660 \sigma_\delta'.
\end{equation}

Using the transformations $\gamma_\alpha = r \cos\theta$ and $\gamma_\delta = (a_\delta'/a_\alpha') r \sin\theta$, we can show that the probability contained within an equiprobability ellipse that passes through a point with separations $\gamma_\alpha$ and $\gamma_\delta$ is
\begin{equation}
P(\gamma_\alpha, \gamma_\delta) = 1 - e^{-\left[(\gamma_\alpha/a_\alpha')^2+(\gamma_\delta/a_\delta')^2\right]/2}.
\label{equation:balrog-enclosed}
\end{equation}

\ADD{In our analysis below, we will use both the two-dimensional images (including systematic errors) and equation~(\ref{equation:balrog-enclosed}). Both give very similar results, which validates our approach to incorporating the systematic errors into the BALROG fits.}

\section{Samples}
\label{section:samples}

We evaluate the two algorithms using a full sample of 54 GRBs detected by both {\FermiGBM} and {\SwiftBAT} and with published positions from BAT, RoboBA, and BALROG and using a bright \CHANGE{sample}{subsample} of 27 GRBs. In this section, we describe the construction of the samples.

\subsection{Full Sample}

We first generated a list of all of the GRBs detected by GBM between 2019 September 14 UTC and 2023 November 07 UTC. The start date was chosen to be when version 41731 of the RoboBA ground software started to be used (for trigger 590141799 corresponding to GRB 190914.345). The end date has no particular external significance but was when we began the final analysis for publication. This interval excludes the GRBs used to calibrate the two algorithms. To create the list, we examined the GCN Notices distributed by the GBM team and ignored triggers that did not have “GRB” as the \verb|MOST_LIKELY| classification value in the latest notice (e.g., GRB 220826497 was initially classified as most likely to be distant particles, but subsequently reclassified as most likely to be a GRB). This gave a list of 984 GRBs.

We then generated a list of all GRBs detected by BAT in the same interval from the “{\Swift} Trigger and Burst Real-Time Information” table on the GCN website\footnote{\url{https://gcn.gsfc.nasa.gov/swift_grbs.html}}. This gave a list of 398 GRBs.

We matched the two lists under the assumption that GRBs that have trigger times within 100 seconds are the same. This gave a list of 74 GRBs detected by both GBM and BAT. Since these events were detected by both satellites, there is a strong likelihood that they are real astrophysical bursts.

We then matched the GBM bursts with the BALROG positions in the “GBM-Locations” catalog on the MPE website\footnote {\url{https://grb.mpe.mpg.de/grb_overview/}}. We found positions for 54 of the 74 GRBs and found one other (GRB~200427768) for which the BALROG analysis was noted as having failed. We do not know why there are no BALROG positions for the other 19 GRBs. While some are faint, others are bright enough to have small RoboBA statistical uncertainty. We do not use these 20 GRBs in our analysis, but only the 54 for which published positions from both RoboBA and BALROG are available. These 54 GRBs form our full sample.

Table~\ref{table:sample} shows the list of 74 GRBs detected by both GBM and BAT. The first column shows the GBM GRB name (year, month, day, and thousandths of a day) and the next two columns show the BAT and GBM trigger numbers. After that, the table shows the positions from BAT, RoboBA, and BALROG. The uncertainties on the BAT positions are typically 3~arcmin in radius (90\% probability) and are negligible compared to the uncertainties of at least 1 deg in the position estimates from both RoboBA and BALROG. Therefore, we take the BAT positions to be the true position $\alpha$ and $\delta$. For RoboBA we show the version of the software, the estimated position $\hat\alpha$ and $\hat\delta$, and the statistical uncertainty $\sigma_\mathrm{stat}$ defined in section~\ref{section:roboba-definitions}. For BALROG, we show the estimated position $\hat\alpha$ and $\hat\delta$, the statistical uncertainties $\sigma_\alpha$ and $\sigma_\beta$, and the systematic uncertainty $\sigma_\mathrm{sys}$, defined in section~\ref{section:balrog-definitions}. 

Most of the RoboBA positions are ground positions produced by version 41731 of the software, but a few are flight positions produced by version 3 or ground positions produced by versions 415 or 4173.

Table~\ref{table:derived} shows derived quantities for the full sample of 54 GRBs used in the analysis. In particular, it shows the error $\gamma$ in deg between the positions estimated by RoboBA and BALROG and the true position determined by BAT.

\subsection{Bright \CHANGE{Sample}{Subsample}}

Our full sample includes both bright and faint GRBs and as such is in some ways unfair to BALROG, which is optimized to give improved positions of the brightest GRBs. For example, the current implement of BALROG uses only 10 seconds of data around the peak to avoid smearing due to the motion of the spacecraft \citep{berlato2019improved}, whereas our understanding is that RoboBA uses a longer interval and so might be expected to give better results for fainter GRBs dominated by statistical errors.

The question of whether a GRB is bright in this sense is not completely clearly defined. One might consider only GRBs with $T_{90}$ of no more than 10 seconds, so that BALROG sees almost all of the flux, or consider the peak flux. However, we have adopted an empirical approach; essentially, we ask BALROG if it considers a GRB to be well-localized or not\ADD{ according to the statistical uncertainty, which is indirectly related to the number of photons analysed by BALROG}. For each GRB, we determine the equivalent BALROG circular statistical uncertainty $\sigma$ using
\begin{equation}
\sigma^2 \equiv (\sigma_\alpha\cos\delta)(\sigma_\delta).
\label{equation:sigma}
\end{equation}
Table~\ref{table:derived} shows $\sigma$ in deg for each GRB.
We then define a bright \CHANGE{sample}{subsample} of 27 GRBs that includes only those GRBs with equivalent circular uncertainties smaller than the median of $5.7~\deg$.  

\subsection{BALROG Map \CHANGE{Sample}{Subsample}}

The BALROG “GBM-Locations” catalog on the MPE website also contains FITS HEALPIX images containing the probability distribution on the sky. Unfortunately, we were only able to find images for the 25 GRBs from GRB~210211363 and later. We refer to these 25 as the BALROG map \CHANGE{sample}{subsample}. Furthermore, 12 are also in our bright \CHANGE{sample}{subsample} and so form the bright BALROG map \CHANGE{sample}{subsample}.

\section{Analysis}
\label{section:analysis}

\subsection{Distribution of Errors}

We first consider the distribution of the errors in the position. For each GRB in our sample, we  determined the error as the angular distance $\gamma$ between the BAT position, which is assumed to be the true position, and the positions estimated by the RoboBA and BALROG algorithms.

\begin{figure}
\includegraphics[width=\linewidth]{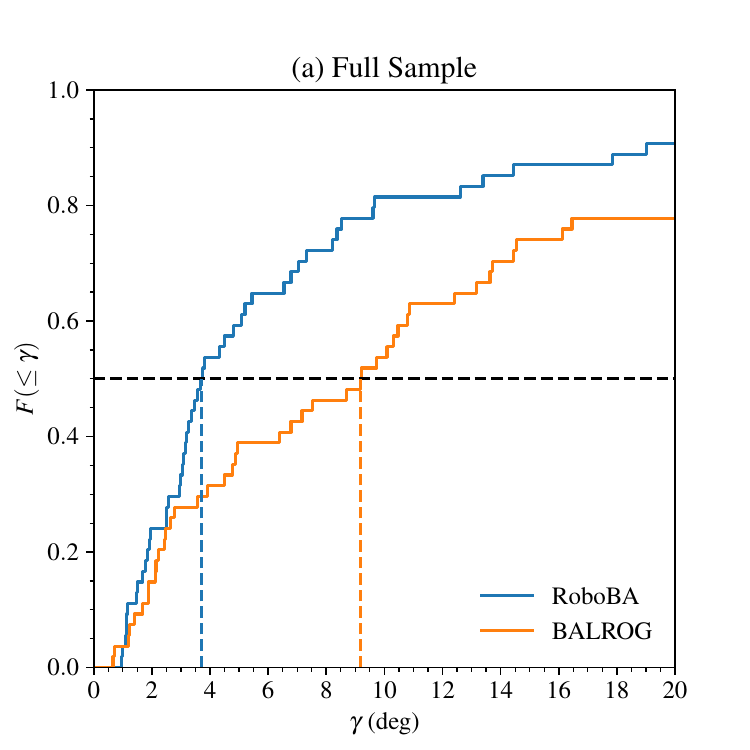}
\includegraphics[width=\linewidth]{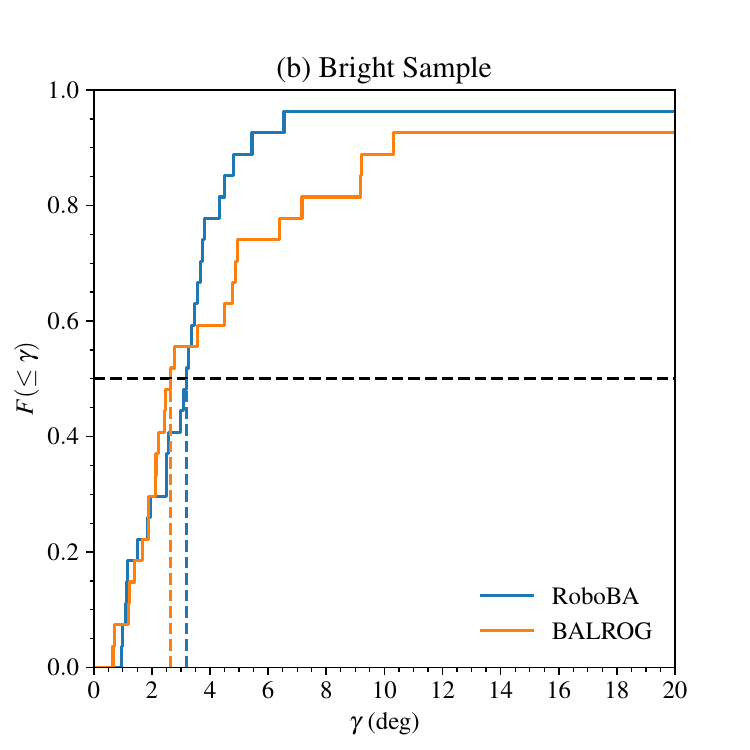}
\caption{The cumulative distribution of position errors $\gamma$ for the RoboBA and BALROG algorithms for the GRBs in (a) the full sample (above) and (b) the bright \CHANGE{sample}{subsample} (below). The median errors, shown with dashed lines, are 3.7 and $3.2~\deg$ for RoboBA and 9.2 and $2.6~\deg$ for BALROG.}
\label{figure:error-distribution}
\end{figure}

Figure~\ref{figure:error-distribution}\ADD{a} shows the cumulative distribution of errors for both algorithms and for the full sample of 54~GRBs. We see that the two algorithms give very similar results for the roughly 25\% of GRBs with errors of up to about 3~degrees, but after that, the RoboBA algorithm has smaller errors than the BALROG algorithm. The median errors, shown with dashed lines in Figure~\ref{figure:error-distribution}a, are $3.7~\deg$ for RoboBA and $9.2~\deg$ for BALROG.

Figure~\ref{figure:error-distribution}b shows the cumulative distribution of errors for both algorithms and for the bright \CHANGE{sample}{subsample}. The results are now quite different. \CHANGE{BALROG initially gives more precise positions than RoboBA, but the crossing point is between 50\% and 60\% of the bright \CHANGE{sample}{subsample} (i.e., between 25\% and 30\% of the full sample), and beyond this, the errors from RoboBA are noticeably better than those of BALROG.}{The two algorithms give very similar results for the better-localized half of the subsample.}  The median errors, shown with dashed lines in Figure~\ref{figure:error-distribution}b, are $3.2~\deg$ for RoboBA and $2.6~\deg$ for BALROG.

The distributions suggest that BALROG is performing similarly or perhaps slightly better than RoboBA for the roughly 25\% of brightest and best-localized GRBs. For the fainter and more poorly localized GRBs, RoboBA gives positions with smaller errors. This change in behavior is consistent with the stated optimization of BALROG for brighter GRBs \citep{berlato2019improved}. We considered carrying out statistical tests on these samples to quantify these statements further. However, with so few GRBs in our samples, we are susceptible to large statistical fluctuations and also to somewhat arbitrary decisions (e.g., our choice of the median uncertainty to define the bright \CHANGE{sample}{subsample}, rather than a smaller or larger percentile\ADD{, such as the bright \CHANGE{sample}{subsample} being those with uncertainties in the lowest 25\%}).

One feature that jumps out at us is that even in the bright \CHANGE{sample}{subsample} there are two BALROG positions (GRBs 220118764 and 220714582) and one RoboBA position (GRB 191011192) with errors of more than 60~deg. This gives some idea of the difficulties both groups have faced in finding a robust algorithm.

\subsection{Distribution of Enclosed Probabilities}

\begin{figure}
\includegraphics[width=\linewidth]{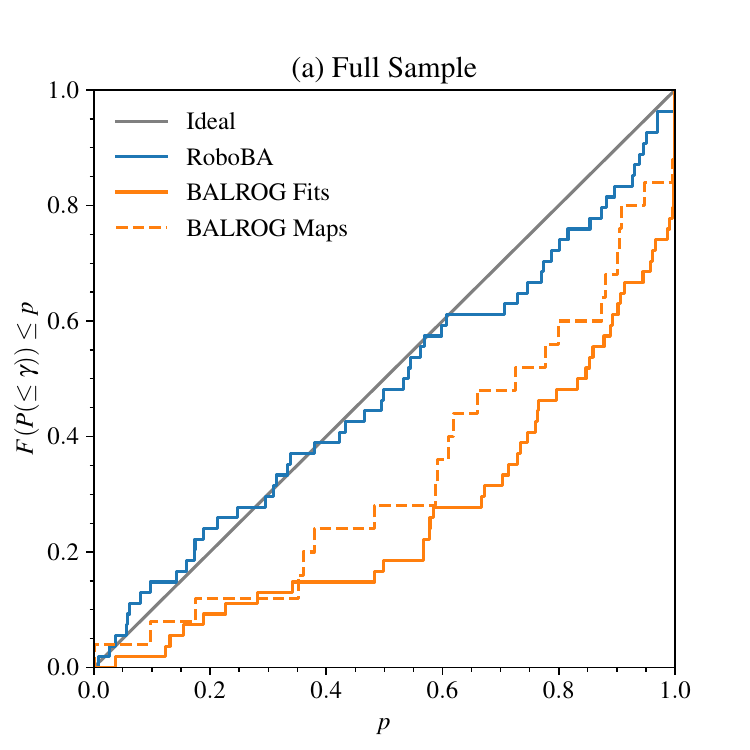}
\includegraphics[width=\linewidth]{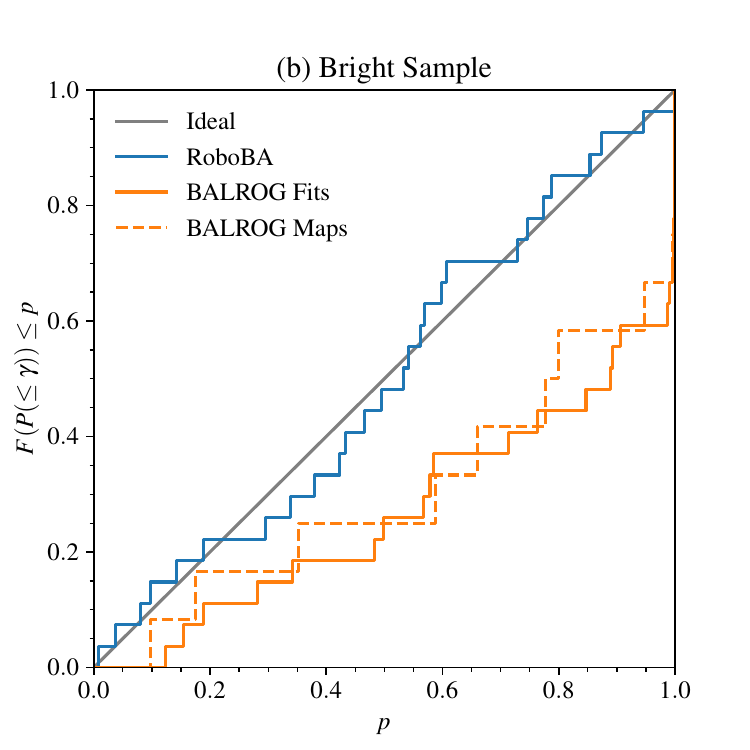}
\caption{The cumulative distribution of enclosed probabilities for the RoboBA and BALROG algorithms for the GRBs in (a) the full sample (above) and (b) the bright \CHANGE{sample}{subsample} (below). For BALROG, the solid lines are determined from the Gaussian fits and the dashed lines are determined directly from the probability maps.}
\label{figure:probability-distribution}
\end{figure}

In addition to positions, both RoboBA and BALROG distribute estimates of the uncertainties in the position. It is important to understand how well these estimates reflect reality. In an ideal case, the cumulative distribution of probability enclosed within the observed offset $\gamma$ should be a straight line from 0 to 1. In the case of RoboBA, the enclosed probability is given by equation (\ref{equation:roboba-enclosed}). In the case of BALROG, we determined it in two ways. For the full sample, we used equation (\ref{equation:balrog-enclosed}) to determine the probability within the equiprobability ellipse passing through the true position. For the map \CHANGE{sample}{subsample}, we directly summed the probability in the maps (the ones that explicitly include the systematic error) within a circle centered on the estimated position and whose radius was the angular distance to the true position. Table~\ref{table:derived} gives these three probabilities.

The cumulative distributions of enclosed probability are shown in Figure~\ref{figure:probability-distribution}a for the whole sample and Figure \ref{figure:probability-distribution}b for the bright \CHANGE{sample}{subsample}. For BALROG, we show two determinations: the solid lines are determined from the Gaussian fits and the dashed lines are determined directly from the probability maps. 

We see that the line for RoboBA is quite close to the ideal case, both for the full sample and the bright \CHANGE{sample}{subsample}. This suggests that the uncertainties are accurately given by RoboBA.
On the other hand, the lines for BALROG are dramatically below the ideal for both samples and for both probabilities determined from the fits and from the maps. This strongly suggests the true uncertainties in the position are underestimated by BALROG. 

In the full sample, the fraction of GRB errors within the estimated $1\sigma$ ($P = 0.6827$) uncertainty region is 61\% for RoboBA, 31\% for BALROG with probabilities determined from the fits, and 48\% for BALROG with probabilities determined from the maps. For the bright \CHANGE{sample}{subsample}, the corresponding percentages are 70\% for RoboBA and 37\% and 42\% for BALROG.

Again, we need to remind ourselves that BALROG was optimized for bright GRBs \citep{berlato2019improved}. Therefore, we should not be too demanding of its performance with the full sample, which includes both bright and faint GRBs. However, it is somewhat surprising that the cumulative distribution for the bright \CHANGE{sample}{subsample} differs so much from the ideal case. 

To quantify the magnitude of the effect, we found that if we artificially increased the BALROG uncertainties from the fits by a factor of two, the percentage within the estimated $1\sigma$ ($P = 0.6827$) uncertainty region rose from 37\% to 63\%. Thus, it appears that the uncertainties generated by BALROG are underestimated by roughly a factor of two.

\section{Application to DDOTI}
\label{section:ddoti}

\begin{figure*}
\centering
\includegraphics[width=0.45\linewidth]{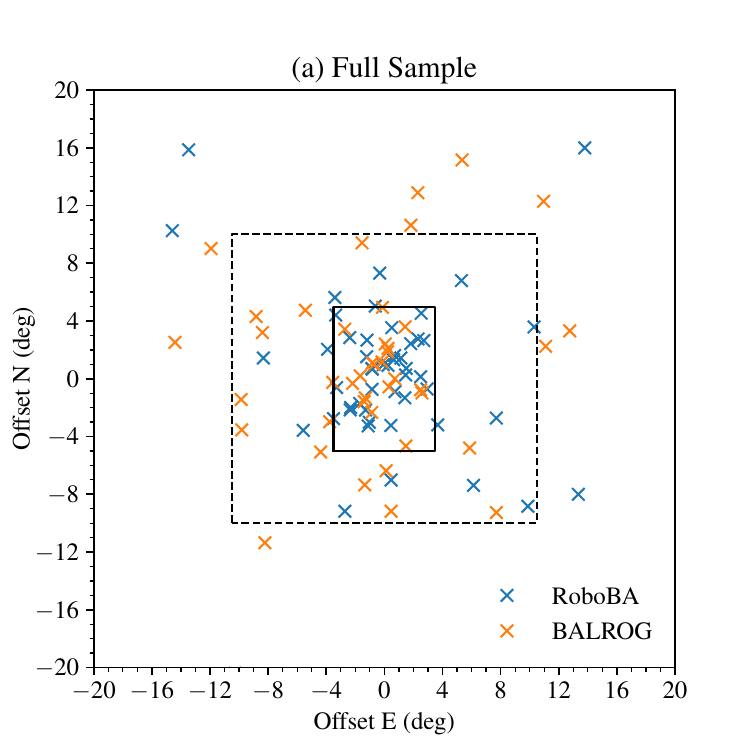}
\includegraphics[width=0.45\linewidth]{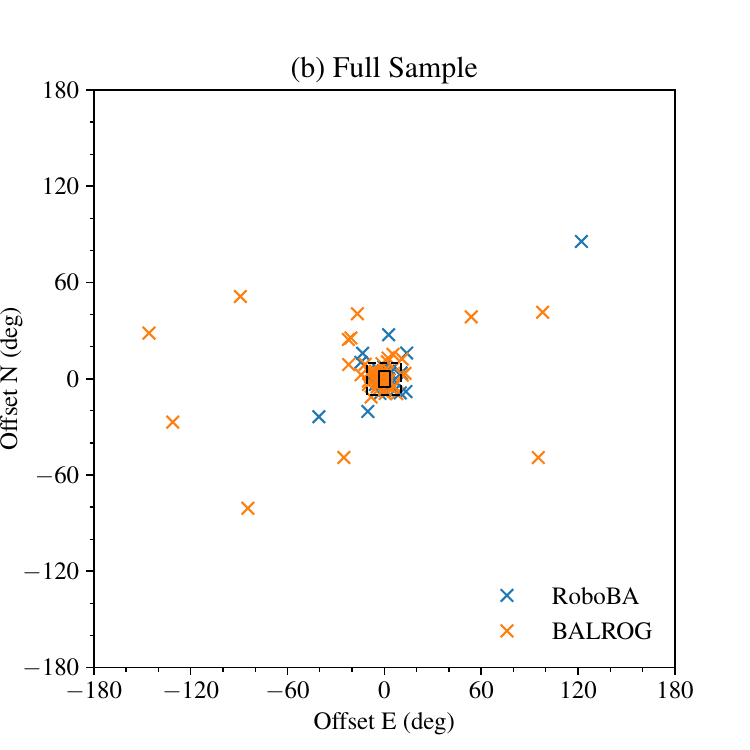}

\includegraphics[width=0.45\linewidth]{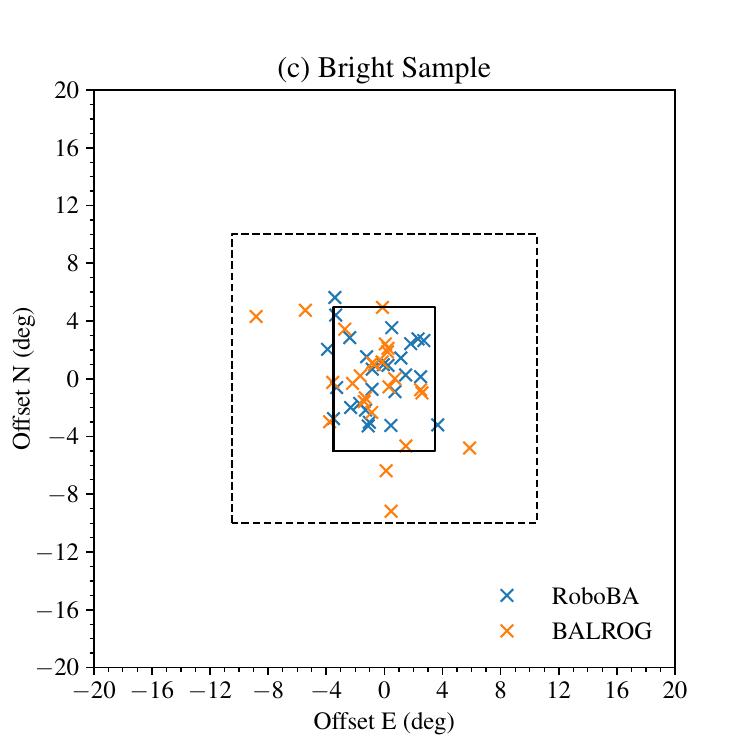}
\includegraphics[width=0.45\linewidth]{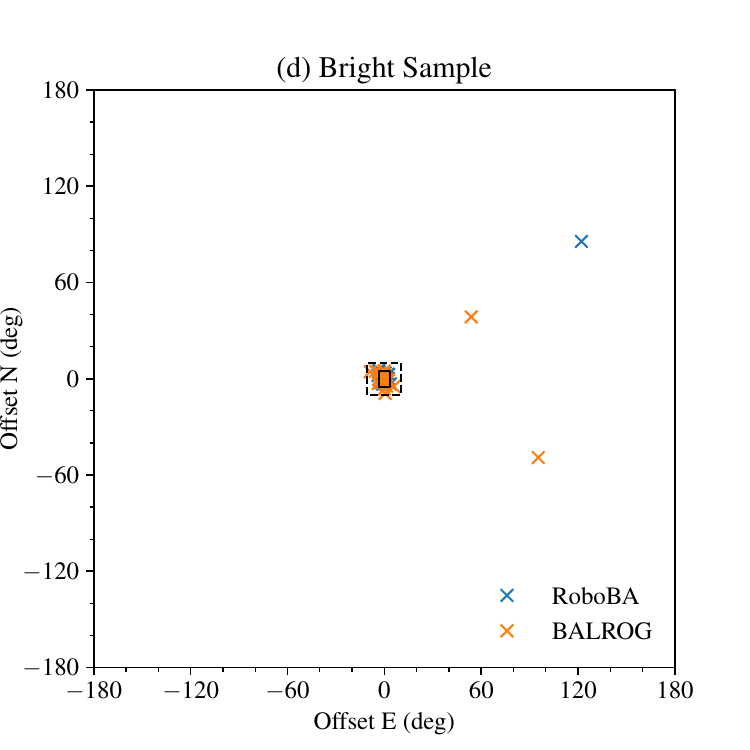}
\caption{The errors in the positions shown relative to the DDOTI fields. The solid black rectangle represents one DDOTI field ($7 \times 10$~deg). The dashed black rectangle represents a mosaic of $3\times2$ DDOTI fields ($21 \times 20$~deg). The crosses show the errors in the RoboBA and BALROG positions. The upper panels are for the full sample and the lower panels are for the bright \CHANGE{sample}{subsample}. The left panels zoom in on the DDOTI field. The right panels show the tail of outliers.}
\label{figure:error-scatter}
\end{figure*}

DDOTI\footnote{\url{http://ddoti.astroscu.unam.mx/}} is a wide-field imager installed at the Observatorio Astronómico Nacional on the Sierra de San Pedro Mártir in Baja California, Mexico \citep{watson2016ddoti}. It consists of six 28 cm telescopes with prime-focus CCDs mounted on a common equatorial mount and offers an instantaneous field of view of approximately 70~{\sqdeg} (7~deg E-W and 10~deg N-S) with a $10\sigma$ limited magnitude in 60 seconds of approximately $r \approx 18.7$ in dark time and $r \approx 18.0$ in bright time. Its main science goals are the localization and follow-up of optical transients associated with GRBs detected by {\FermiGBM} and gravitational-wave events \citep{watson2020limits,thakur2020search,becerra2021ddoti,dichiara2021ddoti}.

In its follow-up of GRBs detected by {\FermiGBM}, DDOTI is a consumer of the estimated positions and uncertainties produced by the RoboBA and BALROG algorithms. At a practical level, we need to understand whether we should point the telescope using the RoboBA or BALROG positions and whether we should observe a single field or a larger mosaic. The second issue involves a trade-off between coverage and depth. DDOTI is typically background-limited and the optical transients fade rapidly, so if we observe at $N$ pointings we increase our effective field to about $70N${\sqdeg} but reduce our sensitivity by about $1.25\log N$ magnitudes. For mosaics with $1\times1$, $2\times1$, $2\times2$, or $3\times2$ pointings, we have fields of $7 \times 10$, $14 \times 10$, $14 \times 20$~deg, and $21 \times 20$~deg and corresponding losses of sensitivity of about 0.0, 0.4, 0.8, and 1.0 magnitudes.

To address these questions, we have simulated observations of our GRB samples with DDOTI. We assume that we center our single-field or mosaic either to the RoboBA or BALROG position, and then ask whether the BAT position would fall within a mosaic with a given size. Table~\ref{table:ddoti} and Figure~\ref{figure:error-scatter} shows the results. We see that centering the mosaic on the RoboBA position gives a higher fraction of GRBs within the field than centering the mosaic on the BALROG position. For example, if we choose to observe only one field, then pointing to the RoboBA position would give coverage at the true position of 57\% of the full sample, whereas the same for the BALROG position would give only 35\%. The corresponding percentages for the bright \CHANGE{sample}{subsample} are 81\% for RoboBA and 67\% for BALROG.

Considering the high fraction of GRBs within one field and the significant sensitivity loss for observing more than one pointing, our strategy for localizing GBM GRBs with DDOTI is to observe only a single pointing centered on the RoboBA position.

\section{\CHANGE{Discussion}{Conclusions}}
\label{section:discussion}

We have evaluated the precision of the estimates of GRB positions and the precision of the associated uncertainties for a sample of 54 GRBs detected by both {\SwiftBAT} and {\FermiGBM} and for which positions are available with \CHANGE{recenter}{recent} versions of the RoboBA and BALROG algorithms. 

We find that RoboBA and BALROG offer very similar results for approximately 25\% of GRBs of bright GRBs with errors up to about 3~deg, RoboBA gives smaller errors beyond this range. This result was not entirely surprising, since BALROG was optimized for bright GRBs in which the systematic error can dominate \citep{berlato2019improved}, and some of the improvements in RoboBA seem to have been stimulated by the earlier advances in BALROG. The approach taken by RoboBA minimizes the statistical error, which dominates for faint GRBs.

On the other hand, we find that while the uncertainties estimated by RoboBA correspond closely to the observed uncertainties, those produced by BALROG seem to underestimate the true uncertainties significantly, perhaps by a factor of two. This was an unexpected result. The uncertainties in the position of the GRB are important for defining the limits of the search area, both when determining the observing strategy and during subsequent analysis.

One of the advantages of BALROG, demonstrated by \cite{burgess2018}, is that it makes explicit the dependency between the source position and spectrum by solving for these simultaneously. Our results do not contradict this finding, as we focus only on the position. This suggests that it would be worthwhile to understand why the estimates of the positional uncertainties produced by BALROG seem to be underestimated, both to have confidence in its fits to spectra and to continue to be able to work with the correlations in the fits.

RoboBA emerges as a better choice for determining the observing strategy of our wide-field imager DDOTI. The combination of the wide field of DDOTI (70~\sqdeg) and the improvements in GBM positions over the last few years suggest that in a single pointing, DDOTI can image 57\% of GBM GRBs and 81\% of the brighter half. This argues against mosaicing multiple fields. Obviously, the determination of how many fields should be observed will be different for other instruments with smaller fields.

It is clear that there is still work to be done in this field. RoboBA is producing excellent positions, both for bright and faint GRBs, but there are nagging worries about its treatment of the dependence of the position on the spectrum \citep{burgess2018,berlato2019improved}. The current implementation of BALROG is producing excellent positions for bright GRBs, but seems to be underestimating the uncertainties. As consumers of the results of these algorithms, we look forward to future improvements in both.

\section*{Acknowledgements}

\ADD{We are grateful to an anonymous referee for suggestions that improved the presentation of our results.} We are grateful to Jochen Greiner and Tobias Preis for useful comments on an early draft of this paper. KOCL acknowledges support from a CONAHCyT fellowship. RLB acknowledges support from a CONAHCyT postdoctoral fellowship. We are also grateful for support from UNAM DGAPA/PAPIIT projects IN105921 and IN109224 and CONAHCyT project 277901.  

\section*{\CHANGE{Research Data Policy}{Data Availability}}
The data underlying this article will be shared on reasonable request to the corresponding author.

\bibliographystyle{mnras}
\bibliography{references} 

\begin{landscape}

\begin{table}
\caption{The Sample of GRBs. Positions and uncertainties are in deg.}
\label{table:sample}

\begin{tabular}{rrrrrrrrrrrrrrrrrr}
\hline
\multicolumn{1}{c}{GRB}&
\multicolumn{1}{c}{BAT}&
\multicolumn{1}{c}{GBM}&
\multicolumn{2}{c}{BAT}&
&
\multicolumn{4}{c}{RoboBA}&
&
\multicolumn{5}{c}{BALROG}\\
\cline{4-5}
\cline{7-10}
\cline{12-16}
&
\multicolumn{1}{c}{Trigger}&
\multicolumn{1}{c}{Trigger}&
\\
&
&
&
\multicolumn{1}{c}{$\alpha$}&
\multicolumn{1}{c}{$\delta$}&
&
\multicolumn{1}{c}{Version}&
\multicolumn{1}{c}{$\hat\alpha$}&
\multicolumn{1}{c}{$\hat\delta$}&
\multicolumn{1}{c}{$\sigma_\mathrm{stat}$}&
&
\multicolumn{1}{c}{$\hat\alpha$}&
\multicolumn{1}{c}{$\hat\delta$}&
\multicolumn{1}{c}{$\sigma_\alpha$}&
\multicolumn{1}{c}{$\sigma_\delta$}&
\multicolumn{1}{c}{$\sigma_\mathrm{sys}$}
\\
\hline
191011192 &  928924 & 592461363 & $ 44.72$ & $-27.84$ && 41731 & $207.94$ & $+57.77$ & $ 4.04$ && $  46.4$ & $ -32.5$ & $  4.6$  & $  5.6$  & $2.0$    \\
191031025 &  932435 & 594175000 & $233.47$ & $ +6.14$ && 41731 & $247.34$ & $+22.13$ & $12.95$ && $ 223.6$ & $  +2.6$ & $  6.1$  & $  7.4$  & $1.0$    \\
191031780 &  932595 & 594240201 & $115.86$ & $-62.34$ && 41731 & $120.85$ & $-59.58$ & $ 1.63$ && $ 110.0$ & $ -58.9$ & $  2.3$  & $  1.0$  & $2.0$    \\
191031891 &  932608 & 594249816 & $283.27$ & $+47.64$ && 41731 & $282.80$ & $+54.96$ & $ 2.35$ && $  97.4$ & $ -33.1$ & $ 78.4$  & $ 40.8$  & $2.0$    \\
191227069 &  946344 & 599103569 & $319.17$ & $-16.70$ && 41731 & $319.70$ & $-13.17$ & $ 1.00$ && $ 318.3$ & $ -15.6$ & $  1.0$  & $  0.6$  & $1.0$    \\
200107810 &  948219 & 600117987 & $107.07$ & $-83.72$ && 41731 & $356.88$ & $-80.13$ & $ 8.29$ && $  93.0$ & $ -74.3$ & $ 14.7$  & $  9.5$  & $2.0$    \\
200109074 &  948361 & 600227156 & $307.17$ & $+53.00$ && 41731 & $284.70$ & $+68.86$ & $10.41$ && $ 293.2$ & $ +56.2$ & $ 14.0$  & $  9.4$  & $1.0$    \\
200215611 &  956639 & 603470376 & $ 34.12$ & $+12.78$ && 41731 & $ 30.53$ & $+10.02$ & $ 4.55$ && $  32.7$ & $ +11.2$ & $  4.5$  & $  5.9$  & $2.0$    \\
200216564 &  956824 & 603552759 & $160.46$ & $+19.45$ && 41731 & $159.17$ & $+20.98$ & $ 6.54$ && $ 154.7$ & $ +24.2$ & $  3.6$  & $  3.1$  & $1.0$    \\
200219317 &  957271 & 603790614 & $342.64$ & $-59.11$ && 41731 & $344.08$ & $-60.00$ & $13.92$ && $ 325.4$ & $ -54.8$ & $  7.9$  & $  6.1$  & $1.0$    \\
200227306 &  958592 & 604480813 & $ 56.44$ & $ +9.49$ && 41731 & $ 56.89$ & $ +6.26$ & $ 1.00$ && $  56.5$ & $ +11.9$ & $  1.5$  & $  1.9$  & $2.0$    \\
200228469 &  958733 & 604581288 & $252.03$ & $+16.96$ && 41731 & $258.44$ & $ +9.58$ & $16.72$ && $ 228.9$ & $ +25.8$ & $ 16.6$  & $  9.8$  & $2.0$    \\
200303107 &  959431 & 604895668 & $212.70$ & $+51.36$ && 41731 & $207.44$ & $+50.75$ & $ 3.80$ && $ 216.7$ & $ +50.6$ & $  2.4$  & $  1.3$  & $2.0$    \\
200411187 &  965784 & 608272147 & $ 47.66$ & $-52.31$ && 41731 & $ 69.58$ & $-60.32$ & $24.08$ && $  68.6$ & $ -49.0$ & $ 20.1$  & $ 20.9$  & $1.0$    \\
200427768 &  968211 & 609704785 & $293.74$ & $+21.89$ &&  4173 & $293.40$ & $+17.94$ & $ 7.47$ && ---      & ---      & ---      & ---      \\
200528436 &  974827 & 612354449 & $176.64$ & $+58.15$ && 41731 & $169.22$ & $+60.19$ & $ 1.00$ && $ 169.9$ & $ +57.9$ & $  1.2$  & $  0.6$  & $1.0$    \\
200529039 &  974942 & 612406604 & $238.75$ & $-11.06$ && 41731 & $239.23$ & $-18.07$ & $ 8.53$ && $ 244.2$ & $  +4.1$ & $  9.6$  & $ 11.2$  & $1.0$    \\
200630076 &  980210 & 615174580 & $ 91.38$ & $-60.79$ && 41731 & $ 86.56$ & $-62.95$ & $ 7.50$ && $  96.1$ & $ -47.9$ & $  8.9$  & $  8.1$  & $2.0$    \\
200711461 &  981957 & 616158277 & $285.98$ & $ -0.14$ && 41731 & $287.11$ & $ +1.30$ & $ 1.59$ && $ 286.3$ & $  -0.7$ & $  1.0$  & $  1.2$  & $1.0$    \\
200716957 &  982707 & 616633066 & $196.01$ & $+29.63$ && 41731 & $192.09$ & $+35.26$ & $ 2.33$ && $ 194.5$ & $ +28.3$ & $  1.6$  & $  1.3$  & $2.0$    \\
200801842 &  985320 & 618005512 & $281.63$ & $ -2.99$ &&     3 & $267.02$ & $ +7.27$ & $30.82$ && $ 292.6$ & $  +9.3$ & $ 45.6$  & $ 30.3$  & $1.0$    \\
200903031 &  994389 & 620786664 & $164.31$ & $+50.50$ && 41731 & $159.04$ & $+54.91$ & $ 3.56$ && $ 161.7$ & $ +50.7$ & $  3.1$  & $  2.3$  & $1.0$    \\
200906550 &  994856 & 621090718 & $272.29$ & $+67.85$ && 41731 & $270.00$ & $+68.56$ & $ 4.08$ && $ 268.7$ & $ +60.5$ & $ 17.8$  & $  4.8$  & $2.0$    \\
201001416 &  998344 & 623239145 & $110.07$ & $ -2.21$ && 41731 & $111.35$ & $ +6.45$ & $ 3.93$ && ---      & ---      & ---      & ---      & ---      \\
201006054 &  998907 & 623639877 & $ 61.88$ & $+65.15$ &&   415 & $ 37.27$ & $+44.74$ & $17.55$ && $  88.5$ & $ +67.4$ & $ 21.3$  & $  7.2$  & $2.0$    \\
201017407 & 1000613 & 624620797 & $ 36.63$ & $+66.67$ && 41731 & $ 21.74$ & $+63.51$ & $ 7.27$ && ---      & ---      & ---      & ---      & ---      \\
201021852 & 1001130 & 625004846 & $ 12.55$ & $-55.84$ &&     3 & $ 47.05$ & $-23.60$ & $29.35$ && ---      & ---      & ---      & ---      & ---      \\
201029847 & 1003002 & 625695596 & $229.60$ & $+44.46$ && 41731 & $217.92$ & $+45.90$ & $ 6.33$ && $ 240.4$ & $ +35.2$ & $ 45.0$  & $ 33.0$  & $2.0$    \\
201105099 & 1004219 & 626235728 & $277.68$ & $ -6.75$ &&  4173 & $277.05$ & $ -1.72$ & $12.09$ && $ 256.8$ & $ +18.7$ & $ 15.1$  & $ 14.0$  & $1.0$    \\
201216963 & 1013243 & 629852850 & $ 16.36$ & $+16.54$ && 41731 & $ 17.88$ & $+16.80$ & $ 1.00$ && $  16.2$ & $ +17.7$ & $  0.5$  & $  0.9$  & $2.0$    \\
210102861 & 1015728 & 631312759 & $235.74$ & $-37.22$ && 41731 & $239.15$ & $-34.58$ & $ 3.44$ && $ 231.0$ & $ -40.2$ & $  2.2$  & $  2.6$  & $1.0$    \\
210104477 & 1015873 & 631452424 & $103.70$ & $+64.66$ && 41731 & $107.94$ & $+67.09$ & $ 2.88$ && $ 103.4$ & $ +69.6$ & $  5.8$  & $  1.5$  & $1.0$    \\
210119121 & 1017711 & 632717654 & $282.80$ & $-61.80$ && 41731 & $285.78$ & $-63.12$ & $ 7.81$ && $ 246.9$ & $ -21.3$ & $110.4$  & $ 41.6$  & $2.0$    \\
210211363 & 1032024 & 634725803 & $269.43$ & $-46.30$ && 41731 & $283.75$ & $-55.13$ & $11.78$ && $ 271.5$ & $ -42.7$ & $ 37.8$  & $ 20.9$  & $1.0$    \\
210306162 & 1035994 & 636695642 & $129.97$ & $+60.20$ && 41731 & $139.18$ & $+64.23$ & $ 1.00$ && ---      & ---      & ---      & ---      & ---      \\
210306397 & 1036024 & 636715939 & $331.85$ & $+10.18$ &&  4173 & $337.91$ & $+14.72$ & $ 6.46$ && ---      & ---      & ---      & ---      & ---      \\
210308276 & 1036227 & 636878281 & $ 67.09$ & $+37.41$ && 41731 & $ 63.45$ & $+42.77$ & $ 1.18$ && ---      & ---      & ---      & ---      & ---      \\
210610628 & 1054627 & 645030227 & $204.28$ & $+14.48$ &&   415 & $212.24$ & $+11.76$ & $ 4.94$ && $ 189.4$ & $ +17.0$ & $ 14.4$  & $  9.8$  & $2.0$    \\
210610827 & 1054681 & 645047470 & $243.94$ & $+14.39$ && 41731 & $241.54$ & $+12.40$ & $ 2.79$ && $ 244.2$ & $ +16.5$ & $  0.6$  & $  1.1$  & $1.0$    \\
210618072 & 1056426 & 645673421 & $235.82$ & $+46.04$ && 41731 & $227.77$ & $+42.47$ & $ 7.60$ && $ 221.6$ & $ +44.6$ & $ 12.4$  & $  5.8$  & $2.0$    \\
210712405 & 1059881 & 647775795 & $ 97.34$ & $-35.39$ && 41731 & $112.13$ & $-32.42$ & $ 6.44$ && ---      & ---      & ---      & ---      & ---      \\
210722871 & 1061223 & 648680085 & $ 27.02$ & $ -6.35$ &&   415 & $ 24.69$ & $ -6.97$ & $ 3.10$ && ---      & ---      & ---      & ---      & ---      \\
210723615 & 1061284 & 648744372 & $121.73$ & $-32.89$ &&   415 & $118.89$ & $-30.04$ & $ 2.96$ && $ 120.9$ & $ -31.9$ & $  4.2$  & $  2.7$  & $2.0$    \\
\hline
\end{tabular}
\end{table}
\end{landscape}

\begin{landscape}
\begin{table}
\contcaption

\begin{tabular}{rrrrrrrrrrrrrrrrrr}
\hline
\multicolumn{1}{c}{GRB}&
\multicolumn{1}{c}{BAT}&
\multicolumn{1}{c}{GBM}&
\multicolumn{2}{c}{BAT}&
&
\multicolumn{4}{c}{RoboBA}&
&
\multicolumn{5}{c}{BALROG}\\
\cline{4-5}
\cline{7-10}
\cline{12-16}
&
\multicolumn{1}{c}{Trigger}&
\multicolumn{1}{c}{Trigger}&
\\
&
&
&
\multicolumn{1}{c}{$\alpha$}&
\multicolumn{1}{c}{$\delta$}&
&
\multicolumn{1}{c}{Version}&
\multicolumn{1}{c}{$\hat\alpha$}&
\multicolumn{1}{c}{$\hat\delta$}&
\multicolumn{1}{c}{$\sigma_\mathrm{stat}$}&
&
\multicolumn{1}{c}{$\hat\alpha$}&
\multicolumn{1}{c}{$\hat\delta$}&
\multicolumn{1}{c}{$\sigma_\alpha$}&
\multicolumn{1}{c}{$\sigma_\delta$}&
\multicolumn{1}{c}{$\sigma_\mathrm{sys}$}
\\
\hline
210725158 & 1061511 & 648877628 & $215.36$ & $ -1.18$ &&   415 & $217.90$ & $ +3.36$ & $ 3.46$ && $   1.2$ & $ +27.3$ & $155.5$  & $ 42.6$  & $2.0$    \\
210731931 & 1062336 & 649462872 & $300.31$ & $-28.04$ && 41731 & $302.02$ & $-27.32$ & $ 2.50$ && $ 291.0$ & $ -39.4$ & $ 18.6$  & $ 17.4$  & $1.0$    \\
210824174 & 1070157 & 651471008 & $232.13$ & $+11.13$ && 41731 & $235.13$ & $+10.43$ & $ 5.34$ && $   8.3$ & $ -15.9$ & $122.7$  & $ 41.1$  & $1.0$    \\
211129410 & 1085430 & 659872271 & $274.56$ & $+31.78$ && 41731 & $271.36$ & $+22.60$ & $ 7.28$ && $ 269.4$ & $ +26.7$ & $  7.9$  & $  8.5$  & $1.0$    \\
211211549 & 1088940 & 660921004 & $212.27$ & $+27.88$ && 41731 & $211.31$ & $+27.15$ & $ 1.00$ && $ 215.2$ & $ +26.9$ & $  0.1$  & $  0.1$  & $2.0$    \\
220118764 & 1093742 & 664222840 & $192.27$ & $+22.91$ && 41731 & $191.37$ & $+23.58$ & $ 5.03$ && $ 299.0$ & $ -26.2$ & $  1.6$  & $  4.0$  & $1.0$    \\
220403863 & 1101053 & 670711364 & $191.03$ & $+89.17$ &&  4173 & $154.09$ & $+80.35$ & $ 5.88$ && ---      & ---      & ---      & ---      & ---      \\
220408240 & 1101675 & 671089569 & $202.40$ & $+47.06$ && 41731 & $192.94$ & $+49.67$ & $ 6.53$ && ---      & ---      & ---      & ---      & ---      \\
220501828 & 1104842 & 673127515 & $ 85.58$ & $+14.03$ && 41731 & $ 71.81$ & $+11.50$ & $ 6.26$ && ---      & ---      & ---      & ---      & ---      \\
220521972 & 1107466 & 674868026 & $275.20$ & $+10.38$ && 41731 & $292.57$ & $ +3.85$ & $13.57$ && ---      & ---      & ---      & ---      & ---      \\
220620016 & 1111002 & 677377371 & $299.39$ & $+35.00$ &&   415 & $142.63$ & $+58.69$ & $ 9.23$ && ---      & ---      & ---      & ---      & ---      \\
220711761 & 1115766 & 679256193 & $261.99$ & $+24.67$ && 41731 & $267.83$ & $+31.47$ & $ 4.61$ && $ 264.0$ & $ +35.3$ & $ 11.0$  & $  9.7$  & $2.0$    \\
220714582 & 1116221 & 679499891 & $ 47.07$ & $-19.33$ && 41731 & $ 45.89$ & $-22.60$ & $ 2.83$ && $ 104.4$ & $ +19.2$ & $  3.3$  & $  3.6$  & $1.0$    \\
220715934 & 1116441 & 679616687 & $254.89$ & $-33.60$ && 41731 & $221.71$ & $-36.44$ & $16.52$ && ---      & ---      & ---      & ---      & ---      \\
220826497 & 1121751 & 683207727 & $206.43$ & $-44.04$ && 41731 & $204.77$ & $-41.37$ & $ 6.56$ && $   1.6$ & $  +7.2$ & $131.8$  & $ 46.0$  & $1.0$    \\
220907587 & 1123129 & 684252331 & $268.87$ & $-20.32$ && 41731 & $269.62$ & $-18.70$ & $ 6.86$ && $ 242.1$ & $ -69.4$ & $182.9$  & $ 49.0$  & $2.0$    \\
221016986 & 1129775 & 687656367 & $ 38.95$ & $-34.62$ && 41731 & $ 37.60$ & $-32.06$ & $ 2.22$ && ---      & ---      & ---      & ---      & ---      \\
221201517 & 1142847 & 691590290 & $266.93$ & $-68.26$ && 41731 & $263.44$ & $-70.43$ & $ 1.46$ && $ 267.5$ & $ -66.4$ & $  1.7$  & $  0.9$  & $1.0$    \\
221216473 & 1144698 & 692882483 & $326.02$ & $-34.41$ && 41731 & $276.31$ & $-58.09$ & $ 5.69$ && $ 298.9$ & $  -9.9$ & $ 12.5$  & $ 13.5$  & $1.0$    \\
221226945 & 1145959 & 693787285 & $ 22.92$ & $-41.55$ && 41731 & $ 16.28$ & $-52.97$ & $ 9.78$ && ---      & ---      & ---      & ---      & ---      \\
230217912 & 1154967 & 698363595 & $280.77$ & $-28.86$ && 41731 & $278.80$ & $-27.39$ & $ 1.00$ && ---      & ---      & ---      & ---      & ---      \\
230328621 & 1162001 & 701708092 & $290.99$ & $+80.02$ && 41731 & $301.82$ & $+75.59$ & $ 3.21$ && ---      & ---      & ---      & ---      & ---      \\
230405832 & 1163119 & 702417488 & $271.47$ & $-47.07$ && 41731 & $276.86$ & $-50.27$ & $ 1.58$ && $ 270.2$ & $ -49.4$ & $  3.7$  & $  1.6$  & $1.0$    \\
230506715 & 1167288 & 705085761 & $134.37$ & $+45.13$ && 41731 & $133.37$ & $+42.86$ & $ 1.59$ && ---      & ---      & ---      & ---      & ---      \\
230723488 & 1180410 & 711805358 & $250.38$ & $ -5.33$ && 41731 & $249.34$ & $ -8.34$ & $ 3.59$ && $ 250.5$ & $ -11.7$ & $  1.9$  & $  2.1$  & $2.0$    \\
230805475 & 1183217 & 712927418 & $207.75$ & $+31.18$ && 41731 & $208.49$ & $+32.50$ & $ 9.10$ && $ 193.8$ & $ +40.2$ & $ 13.8$  & $  7.2$  & $1.0$    \\
230818977 & 1186032 & 714094060 & $285.88$ & $+40.88$ && 41731 & $289.18$ & $+41.01$ & $ 2.69$ && $ 286.5$ & $ +31.7$ & $  3.5$  & $  2.5$  & $1.0$    \\
230826814 & 1187463 & 714771169 & $ 83.01$ & $+66.12$ && 41731 & $ 82.83$ & $+67.10$ & $ 2.02$ && $  77.6$ & $ +65.8$ & $  3.8$  & $  3.9$  & $2.0$    \\
230903724 & 1189514 & 715454583 & $  9.93$ & $-40.92$ && 41731 & $ 13.44$ & $-13.46$ & $16.63$ && $ 185.0$ & $  +0.5$ & $118.2$  & $ 49.5$  & $1.0$    \\
231028173 & 1193078 & 720158951 & $214.03$ & $+20.89$ && 41731 & $214.30$ & $+21.82$ & $ 1.00$ && $ 214.8$ & $ +20.9$ & $  0.7$  & $  0.8$  & $1.0$    \\
231104075 & 1194500 & 720755253 & $ 23.79$ & $+83.79$ && 41731 & $  8.04$ & $+82.08$ & $ 1.00$ && $ 327.3$ & $ +79.0$ & $  1.1$  & $  0.1$  & $2.0$    \\

\hline
\end{tabular}
\end{table}
\end{landscape}

\begin{table*}
\caption{Derived Quantities. $\gamma$ and $\sigma$ are in deg.}
\label{table:derived}

\begin{tabular}{rrrrrrrccc}
\hline
\multicolumn{1}{c}{GRB}&
&
\multicolumn{2}{c}{RoboBA}&
&
\multicolumn{5}{c}{BALROG}\\
\cline{3-4}
\cline{6-10}
&
\\
&
&
\multicolumn{1}{c}{$\gamma$}&
\multicolumn{1}{c}{$P$}&
&
\multicolumn{1}{c}{$\gamma$}&
\multicolumn{1}{c}{$\sigma$}&
\multicolumn{1}{c}{Bright?}&
\multicolumn{1}{c}{$P$ from fit}&
\multicolumn{1}{c}{$P$ from map}
\\
\hline
191011192 && 147.8 & 1.000 &&   4.9 &   4.7 & Y & 0.566 & ---   \\
191031025 &&  20.9 & 0.938 &&  10.5 &   6.7 & N & 0.957 & ---   \\
191031780 &&   3.7 & 0.745 &&   4.5 &   1.1 & Y & 0.986 & ---   \\
191031891 &&   7.3 & 0.970 && 164.8 &  51.8 & N & 0.998 & ---   \\
191227069 &&   3.6 & 0.773 &&   1.4 &   0.8 & Y & 0.764 & ---   \\
200107810 &&  13.4 & 0.926 &&   9.7 &   6.1 & N & 0.704 & ---   \\
200109074 &&  19.0 & 0.969 &&   8.7 &   8.6 & N & 0.764 & ---   \\
200215611 &&   4.5 & 0.541 &&   2.1 &   5.1 & Y & 0.155 & ---   \\
200216564 &&   1.9 & 0.081 &&   7.1 &   3.2 & Y & 0.995 & ---   \\
200219317 &&   1.2 & 0.007 &&  10.3 &   5.3 & Y & 0.991 & ---   \\
200227306 &&   3.3 & 0.729 &&   2.4 &   1.7 & Y & 0.584 & ---   \\
200228469 &&   9.7 & 0.309 &&  23.2 &  12.1 & N & 0.965 & ---   \\
200303107 &&   3.4 & 0.433 &&   2.6 &   1.4 & Y & 0.714 & ---   \\
200411187 &&  14.4 & 0.333 &&  13.6 &  16.6 & N & 0.666 & ---   \\
200528436 &&   4.3 & 0.853 &&   3.6 &   0.6 & Y & 1.000 & ---   \\
200529039 &&   7.0 & 0.498 &&  16.1 &  10.4 & N & 0.913 & ---   \\
200630076 &&   3.1 & 0.158 &&  13.2 &   7.0 & N & 0.945 & ---   \\
200711461 &&   1.8 & 0.338 &&   0.6 &   1.1 & Y & 0.189 & ---   \\
200716957 &&   6.5 & 0.946 &&   1.9 &   1.4 & Y & 0.498 & ---   \\
200801842 &&  17.8 & 0.314 &&  16.4 &  36.9 & N & 0.226 & ---   \\
200903031 &&   5.4 & 0.787 &&   1.7 &   2.1 & Y & 0.483 & ---   \\
200906550 &&   1.1 & 0.055 &&   7.5 &   6.5 & N & 0.902 & ---   \\
201006054 &&  24.5 & 0.881 &&  10.9 &   7.7 & N & 0.877 & ---   \\
201029847 &&   8.3 & 0.800 &&  12.4 &  34.8 & N & 0.131 & ---   \\
201105099 &&   5.1 & 0.173 &&  32.7 &  14.2 & N & 0.998 & ---   \\
201216963 &&   1.5 & 0.295 &&   1.2 &   0.7 & Y & 0.280 & ---   \\
210102861 &&   3.8 & 0.562 &&   4.7 &   2.1 & Y & 0.996 & ---   \\
210104477 &&   3.0 & 0.465 &&   4.9 &   1.7 & Y & 1.000 & ---   \\
210119121 &&   1.9 & 0.058 &&  47.4 &  65.4 & N & 0.672 & ---   \\
210211363 &&  12.6 & 0.706 &&   3.9 &  24.1 & N & 0.036 & 0.002 \\
210610628 &&   8.2 & 0.895 &&  14.5 &  11.6 & N & 0.728 & 0.609 \\
210610827 &&   3.1 & 0.495 &&   2.1 &   0.8 & Y & 0.906 & 0.777 \\
210618072 &&   6.8 & 0.544 &&  10.1 &   7.2 & N & 0.759 & 0.618 \\
210723615 &&   3.7 & 0.607 &&   1.2 &   3.1 & Y & 0.124 & 0.098 \\
210725158 &&   5.2 & 0.769 && 138.1 &  76.7 & N & 0.833 & 0.905 \\
210731931 &&   1.7 & 0.212 &&  13.7 &  15.8 & N & 0.577 & 0.482 \\
210824174 &&   3.0 & 0.246 && 137.2 &  69.6 & N & 0.852 & 0.874 \\
211129410 &&   9.6 & 0.815 &&   6.8 &   7.7 & N & 0.567 & 0.360 \\
211211549 &&   1.1 & 0.187 &&   2.8 &   0.1 & Y & 0.889 & 0.659 \\
220118764 &&   1.1 & 0.038 && 114.2 &   2.4 & Y & 1.000 & 1.000 \\
220711761 &&   8.5 & 0.929 &&  10.8 &   9.3 & N & 0.745 & 0.591 \\
220714582 &&   3.4 & 0.568 &&  68.2 &   3.3 & Y & 1.000 & 1.000 \\
220826497 &&   2.9 & 0.174 && 137.3 &  77.6 & N & 0.859 & 0.901 \\
220907587 &&   1.8 & 0.062 &&  51.7 &  56.2 & N & 0.734 & 0.379 \\
221201517 &&   2.5 & 0.532 &&   1.9 &   0.8 & Y & 0.892 & 0.800 \\
221216473 &&  40.4 & 1.000 &&  34.9 &  12.9 & N & 0.999 & 0.907 \\
230405832 &&   4.8 & 0.873 &&   2.5 &   2.0 & Y & 0.847 & 0.587 \\
230723488 &&   3.2 & 0.423 &&   6.4 &   2.0 & Y & 0.996 & 0.947 \\
230805475 &&   1.5 & 0.026 &&  14.4 &   8.7 & N & 0.960 & 0.725 \\
230818977 &&   2.5 & 0.380 &&   9.2 &   2.7 & Y & 1.000 & 0.995 \\
230826814 &&   1.0 & 0.097 &&   2.2 &   2.5 & Y & 0.578 & 0.352 \\
230903724 &&  27.6 & 0.950 && 139.3 &  76.5 & N & 0.797 & 0.879 \\
231028173 &&   1.0 & 0.141 &&   0.7 &   0.7 & Y & 0.341 & 0.174 \\
231104075 &&   2.6 & 0.598 &&   9.2 &   0.1 & Y & 1.000 & 1.000 \\
\hline
\end{tabular}
\end{table*}
\begin{table*}
\caption{Percentage of GRBs in DDOTI Mosaics}
\label{table:ddoti}

\begin{tabular}{ccccccc}
\hline
\multicolumn{1}{c}{Pointings}&
\multicolumn{1}{c}{Size (deg)}&
\multicolumn{2}{c}{Full Sample}&
&
\multicolumn{2}{c}{Bright Sample}\\
\cline{3-4}
\cline{6-7}
&
&
RoboBA&
BALROG&
&
RoboBA&
BALROG\\
\hline
$1\times1$&$\phantom{0}7\times10$&57\%&35\%&&81\%&67\%\\
$2\times1$&$\phantom{}14\times10$&65\%&43\%&&93\%&81\%\\
$2\times2$&$\phantom{}14\times20$&78\%&52\%&&96\%&89\%\\
$3\times2$&$\phantom{}21\times20$&85\%&61\%&&96\%&93\%\\
\hline
\end{tabular}
\end{table*}

\bsp	
\label{lastpage}
\end{document}